\documentclass[aps,showpacs,superscriptadress,preprint]{revtex4}
\usepackage{graphicx}
\usepackage{amsmath}

\begin{document}
\title{Hypernclear in the improved quark mass density- dependent model}

\author{
Chen Wu$^{1}$\footnote{wuchenoffd@gmail.com}, Yu-Gang Ma$^{1}$,
Wei-Liang Qian$^{2}$, and Ru-Keng Su$^{3}$}
 \affiliation{
\small 1. Shanghai Institute of Applied Physics, Chinese Academic of Sciences, Shanghai 201800, China \\
\small 2. Universidade de Ouro Preto, Ouro Preto, 35400-000, Brazil\\
\small 3. Department of Physics, Fudan University, Shanghai 200433, China\\
}

\begin{abstract}

The improved quark mass density- dependent model,
which has been successfully used to describe the properties of both finite
nuclei and bulk nuclear matter, is extended to include the strange quark.
The parameters of the model are determined by the saturation properties of bulk matter.
Then the given parameter set is employed to investigate
both the properties of strange hadronic matter and those of $\Lambda$ hypernuclei.
Bulk strange hadronic matter consisting of nucleons,
$\Lambda$- hyperons and $\Xi$- hyperons is studied under mean-field approximation.
Among others, density dependence of the effective baryon mass,
saturation properties and stability of the physical system are discussed.
For single-$\Lambda$ hypernuclei, single particle energies of $\Lambda$ hyperon is evaluated.
In particular, it is found that the present model produces a small spin-orbit interaction,
which is in agreement with the experimental observations.
The above results show that the present model can consistently describe the properties of strange hadronic matter,
as well as those of single $\Lambda$ hypernuclei within an uniform parameterization.

\end{abstract}


 \maketitle

\section{Introduction}
Exploring nuclear systems with strangeness, especially those with multiple units of strangeness,
has attracted lots of attention from the researchers for many years.
Such systems have many astrophysical and cosmological implications and are indeed interesting by themselves.
For instance, the core of neutron stars may contain a high fraction of hyperons\cite{ns1,ns2,ns3},
resulting in the third family of compact stars which possess a similar mass to a neutron star but smaller radius\cite{ns1}.
The formation of strange matter in relativistic heavy ion collisions has also been explored.
Recently, the observation of an antimatter hypernucleus $^3_{\bar\Lambda}\bar{H}$
in relativistic heavy-ion collisions was reported by STAR Collaboration\cite{science}.

Strange hadronic matter have been investigated extensively by many authors\cite{Schaffner1}-\cite{S.Zhang}.
In the strange hadronic matter, the strange quarks are localized within individual hyperons,
which are resumed to retain their identity in the bound system.
Schaffner et al. \cite{Schaffner2} discussed for the first time a finite system of strange hadronic matter (MEMOs)
including $\Sigma$ and $\Xi$ hyperons.
Strange hadronic matter in bulk was studied first by Glendenning \cite{Glendenning}.
As pointed out by Schaffner and Gal \cite{Schaffner1}, some early works about strange hadronic matter are incomplete,
either discuss $\Lambda$ matter\cite{strangessold1,strangessold2,strangessold3}
or ignoring $\Xi$ hyperons\cite{strangessold4} or arbitrarily constraining the fraction of strangeness\cite{strangessold5}.
The correct calculation should fulfill the requirements of chemical equilibrium.

On the other hand, the study of hypernuclei provides valuable information on hyperon-nucleon and hyperon-hyperon interactions.
Such information is crucial for understanding the properties of multi-strange systems and neutron stars.
The single-$\Lambda$ hypernucleus is one of the most extensively studied hypernuclei
where the $\Lambda$ hyperon is sitting outside of the closed-shell nuclear core.
$\Lambda$ hypernuclear spectroscopy through the $(\pi^+, K^ +)$
reaction indicates that $\Lambda$ is weakly bound in nuclear medium
and its spin-orbit splitting is quite small compared to that of nucleon \cite{Hypernuclei-Exp3}.
Many theoretical efforts to evaluate hypernuclear structures have been devoted
during past decades in models with hadronic degree of freedom\cite{Keil}-\cite{Deformation}.
Calculations have also been successfully performed using quark meson coupling model
where both hyperon and nucleon are viewed as compositions of quarks \cite{QMC-hypernuclei}-\cite{QMC-hypernuclei2}.
In Ref. \cite{QMC-hypernuclei2}, Guichon et al. studied the properties
of hypernuclei using the latest version of the quark-meson coupling model.
In their work, the effect of the medium on the color-hyperfine interaction due to the gluon exchange between quarks was included,
which turned out to significantly affect the medium hadron (in particular, hyperon) mass.

Recently, an improved quark- mass density dependent model (IQMDD) with quark meson coupling mechanism was proposed\cite{Wu1}-\cite{Wu6}.
The model was in part inspired by the quark mass density- dependent (QMDD) model of Fowler, Raha and Weiner \cite{QMDD1},
where density dependence of quark mass was introduced to achieve phenomenological quark confinement.
To form hadronic matter, quarks interact via the exchange of mesons in the same way as suggested
in quark- meson  coupling (QMC) model first introduced by Guichon \cite{QMC1}.
The QMDD model assumes that the masses of u, d and s quarks (and the corresponding antiquarks) satisfy:
\begin{eqnarray}
m_{q} = \frac{B}{3n_{B}}(q = u,d,\bar u,\bar d)   \label{mq1}
\end{eqnarray}
\begin{eqnarray}
m_{s,\bar s } = m_{s0}+\frac{B}{3n_{B}}     \label{mq2}
\end{eqnarray}
where $n_{B}$ is the baryon number density, $m_{s0}$ is the current
mass of the strange quark, and $B$ is the bag constant.
As explained in Refs. \cite{QMDD2}-\cite{QMDD3}, the $\textmd{ansatz}$ Eqs.(\ref{mq1}) and (\ref{mq2})
corresponds to a quark confinement hypothesis and can replace
the MIT bag boundary condition and produces very similar results.

Quark- meson coupling (QMC) is an hybrid model which successfully describes many
physical properties of nuclear matter and nuclei\cite{QMC2}-\cite{QMC3}.
In this model, the nuclear system was described as a
collection of non-overlapping MIT bags in which quarks interact through the exchange of scalar and vector mesons.
The interactions between quarks and mesons are limited within the MIT bag region.
As was pointed in Refs. \cite{Wu1}-\cite{Wu2}, this model has two major shortcomings:
(1) It cannot describe the quark deconfinement phase transition
since the quarks are confined within the MIT bag by hard boundary condition.
Since the latter is a model hypothesis, deconfinement phase transition does not take place naturally.
(2) It is difficult to do nuclear many-body
calculation beyond mean-field approximation(MFA) by means of QMC model,
because one cannot find the free propagators of quarks and mesons easily.
These may be attributed to the reason that the interactions between quarks and mesons are limited within the bag regions.
In short, these two shortcomings stem from the MIT bag constraint.

The introduction of IQMDD model was aiming at circumventing these difficulties.
Instead of the MIT bag, a Friedberg- Lee soliton bag was made use of in QMC mechanism.
Through the nonlinear interaction of $\sigma$-mesons and their coupling with quarks,
the later are automatically confined owning to the non-topological soliton bag solution of the system.
The bag boundary is determined subsequently by the calculated quark density as a function of radius,
rather than by a prior hypothesis.
In the original work of Friedberg and Lee\cite{fl1,fl2,fl3}, the treatment is temperature independent and (baryon) density invariant.
As a result, the bag boundary does not vary in terms of these quantities.
In order to achieve dynamical deconfinement, it is quite intuitive to further introduce temperature and density dependence into the model.
Inspired by QMDD model, we retain the quark mass density dependence.
While at finite temperature, temperature dependence gives rise to dynamical deconfinement phase transition.
The resulting calculations\cite{Wu1}-\cite{Wu6} showed that the IQMDD model successfully describes the properties
of nuclear matter, neutron stars and finite nuclei
meanwhile it provides an intuitive scenario for quark deconfinement phase transition.
In particular, the model gives a reasonable critical temperature of quark deconfinement\cite{Wu4},
and one may carry out many-body calculations beyond MFA in principle.

The main purpose of present work is to extend the above IQMDD model to study multi-$\Lambda$ matter and $\Lambda$ hypernuclei.
Since the original IMQDD model includes only two flavors of quarks.
In order to provide a reasonable description of nuclear system with strange degree of freedom,
it is necessary to incorporate the strange quark into the model(See Eqs.(\ref{mq1}) and (\ref{mq2})).
In hadronic sector, hyperons are consisted of up, down and strange quarks.
Strange mesons consisting of strange quarks will also be introduced.
The coupling constants between quarks and mesons in the model are essentially determined by the bulk properites of nuclear matter.
Once the parameters are given,
they are used not only in the study of bulk strange hadronic matter, but also in that of $\Lambda$ hypernuclei.
In this context, there is no free parameters in the calculations of $\Lambda$ hypernuclei,
the validity of the model is therefore tested by an uniform parameterization.
In this work, the calculations are carried out at zero temperature.

The paper is organized as follows.
In section II, we give the main formulas for strange nuclear matter and hypernuclei in the IQMDD model.
Numerical results and discussions are presented in section III, calculations are carried out under the MFA.
The last section contains a brief summary.

\section{IQMDD model with strange quarks}

\subsection{Bulk strange hadronic matter with strange mesons}

The IQMDD model is extended in this section by including $\Lambda$ and $\Xi$ hyperons in the system.
An additional hyperon-hyperon (Y-Y) interaction is mediated by two strange mesons
$\sigma^*$ and $\phi$ which couple only to strange quarks,  as proposed by Schaffner et al. \cite{strange mesons}.
For simplicity, we assume that $\Xi^-$s appear in the system in the same proportion as $\Xi^0$s.
Since the system we considered is symmetric for nucleons and $\Xi$s, there is no contribution from $\rho$  mesons.
This is similar to the protons and neutrons in symmetric matter.
The details of the IQMDD model can be found in references \cite{Wu1}-\cite{Wu6}.
Here we only beiefly outline the main formulas of the model.

The Lagrangian density of extended IQMDD model for strange hadronic matter reads:
\begin{eqnarray}
\mathcal{L} &=&
\bar{\psi}_{i/j}[i\gamma^{\mu}\partial_\mu-m_{i}+g^i_\sigma\sigma-g^i_\omega\gamma^\mu\omega_\mu
+g^i_{\sigma^*}\sigma^*-g^i_\phi\gamma^\mu\phi_\mu ]\psi_{i/j}
 \nonumber\\&&+\frac{1}{2}\partial_{\mu}\sigma\partial^{\mu}\sigma \hskip 0.0in
-U(\sigma)
-\frac{1}{4}V_{\mu\nu}V^{\mu\nu}+\frac{1}{2}m_{\omega}^2\omega_\mu\omega^\mu
+\frac{1}{2}m_{\phi}^2{\phi_\mu} \cdot {\phi^\mu}
\nonumber\\&&-\frac{1}{4}{G_{\mu\nu}}{G^{\mu\nu}} +
\frac{1}{2}\partial_{\mu}\sigma^*\partial^{\mu}\sigma^* -
\frac{1}{2}m_{\sigma^*}^2 {\sigma^*}^2
\end{eqnarray}
where
\begin{eqnarray}
 U(\sigma) = \frac{1}{2}m_\sigma^2\sigma^2 +
\frac{1}{3}b\sigma^3+ \frac{1}{4}c\sigma^4+ B,
\end{eqnarray}
\begin{eqnarray}
 -B = \frac{m_\sigma^2}{2}
\sigma_v^2+\frac{b}{3}\sigma_v^3+\frac{c}{4}\sigma_v^4,
\end{eqnarray}
\begin{eqnarray}
\sigma_v=\frac{-b}{2c}[1+\sqrt{1-4m_\sigma^2 c/b^2}],
\end{eqnarray}
where $\psi_{i/j}$ is Dirac spinor for the $i$th quark in the $j$th hadron,
and the quark mass $m_i$ (i = u, d, s) is given by Eqs.(\ref{mq1}) and (\ref{mq2}).
$m_\sigma$, $m_{\sigma^*}$, $m_\phi$
and $m_\omega$ are the masses of $\sigma$, $\sigma^*$, $\phi$ and
$\omega$ mesons respectively, $V_{\mu\nu}=\partial_\mu \omega_\nu-
\partial_\nu \omega_\mu$, ${G}_{\mu\nu}=\partial_\mu {\phi}_\nu-
\partial_\nu {\phi}_\mu$, $g_\sigma^{i}$ and $g_\omega^{i}$  are the
coupling constants between the $i$th quark and $\sigma$ meson and
$\omega$ meson. $g_{\sigma^*}^s$ and $g_\phi^{s}$
are the coupling constants between strange
quark and $\sigma^*$ meson and $\phi$ meson respectively.

The equation of motion for the $i$th quark field in the $j$th hadron (p, n,
$\Lambda$, $\Xi$) under MFA is
\begin{eqnarray}
[i \gamma \cdot \partial - (m_i- g_\sigma^i \bar\sigma -
g_{\sigma^*}^i \bar\sigma^*) - \gamma^0 (g_\omega^i \bar\omega+
g_\phi^i \bar\phi)]\psi_{i/j}=0,
\end{eqnarray}
where $\bar\sigma$,  $\bar\sigma^*$, $\bar\omega$ and $\bar\phi$ are
the mean-field values of the $\sigma$, $\sigma^*$, $\omega$ (the
time component)  and  $\phi$ (the time component)  meson fields,
respectively. The effective quark mass $m_i^{*}$ is given by:
\begin{eqnarray}
 m_i^{*}=m_i-g_\sigma^i\bar{\sigma} -g_{\sigma^*}^i\bar{\sigma^*}.
\end{eqnarray}

In nuclear matter, three quarks constitute a Freidberg-Lee soliton bag,
and the effective baryon mass is obtained from the bag energy and it reads:
\begin{eqnarray}
 M_j^* = \sum_i E_i =\sum_i \frac{4}{3}\pi R^3
\frac{\gamma_i}{(2\pi)^3}\int_0^{K_F^i}\sqrt{{m^*_i}^2+k^2}
(\frac{dN_i}{dk})dk, \label{bage}
\end{eqnarray}
where $\gamma_i$ is the quark degeneracy, $K_F^i$ is Fermi energy of quarks.
$dN_i/dk$ is the density of states for various quarks in a spherical cavity.
The expression of $dN_i/dk$ adopted in this paper can be found in Ref. \cite{Wu1}.

The Fermi energy $K_F^i$ of the $i$th quark is given by
\begin{eqnarray}
3= \frac{4}{3}\pi {R_j}^3 n_{B}^j,
\end{eqnarray}
where $n_B^j$ satisfies
\begin{eqnarray}
 n_B^j =\sum_i \frac{ \gamma_i}{(2\pi)^3} \int_0^{K_F^i}(\frac{dN_i}{dk})  dk.
\end{eqnarray}

The bag radius $R_j$ for the $j$th baryon is determined by the equilibrium condition for the bag energy Eq.(\ref{bage}):
\begin{eqnarray}
\frac{\delta M^*_j}{\delta R_j}=0
\end{eqnarray}

The total energy density at the baryon density $\rho_B$ is given by
\begin{eqnarray}
E_{tot}&=&\sum_j \frac{\gamma_j}{(2\pi)^3}
\int_0^{K_{F_j}} \sqrt{{M_j^*}^2+p^2} dp^3 + \frac{1}{2} m_{\sigma^*}^2 \bar{\sigma^*}^2 \nonumber\\
&& +\frac{1}{2}m_\sigma^2 \bar{\sigma}^2 +\frac{1}{3}b
\bar{\sigma}^3 + \frac{1}{4}c \bar{\sigma}^4+ \frac{1}{2}{m_\phi^2
\bar\phi^2} + \frac{1}{2}{m_\omega^2 \bar\omega^2}, \label{etotal}
\end{eqnarray}
where  the spin-isospin degeneracy $\gamma_j = 4$ for nucleons and
$\Xi$s, and  $\gamma_j = 2$ for $\Lambda$s. The total baryon density
$\rho_B$ is the sum of those of nucleons, $\Lambda$s, and $\Xi$s
\begin{eqnarray}
 \rho_B = \rho_N + \rho_\Lambda + \rho_\Xi. \label{drhob}
\end{eqnarray}

The Fermi momentum $k_{F_j}$ is determined by the relations:
\begin{eqnarray}
 \rho_j =\frac{\gamma_j k^3_{F_j}}{6\pi^2}.
\end{eqnarray}

The $\bar\omega$ and $\bar\phi$ fields  are determined by baryon
number conservation, their values are expressed by
\begin{eqnarray}
 \bar\omega=\frac{3g_\omega^q \rho_N + 2g_\omega^q \rho_\Lambda + g_\omega^q
 \rho_\Xi}{m_\omega^2}, (q= u, d)
\end{eqnarray}
and
\begin{eqnarray}
 \bar\phi=\frac{g_\phi^s \rho_\Lambda + 2g_\phi^s \rho_\Xi}{m_\phi^2}
\end{eqnarray}

The scalar mean field $\bar\sigma$ and $\bar\sigma^*$ are determined
by the self-consistent condition:
\begin{eqnarray}
m_\sigma^2 \bar\sigma+b\bar\sigma^2+c\bar\sigma^3 = - \Sigma_j
\frac{\gamma_j}{(2\pi)^3} \int_0^{K_{F_j}}
\frac{M_j^*}{\sqrt{{M_j^*}^2+p^2}} d^3 p (\frac{\partial
M_j^*}{\partial \bar\sigma})_{R_j},
\end{eqnarray}
and
\begin{eqnarray}
m_{\sigma^*}^2 \bar\sigma^* = - \Sigma_j \frac{\gamma_j}{(2\pi)^3}
\int_0^{K_{F_j}} \frac{M_j^*}{\sqrt{{M_j^*}^2+p^2}} d^3 p
(\frac{\partial M_j^*}{\partial \bar\sigma^*})_{R_j}.
\end{eqnarray}

In the system with equal number of protons and neutrons as well as
equal number of $\Xi^0$ and $\Xi^-$, the chemical equilibrium
conditions for the reactions $\Lambda + \Lambda \rightleftharpoons
\Xi^- + p$, and $\Lambda + \Lambda \rightleftharpoons \Xi^0 + n$
read \cite{Schaffner1}
\begin{eqnarray}
2\mu_\Lambda = \mu_N + \mu_\Xi \label{cheq}
\end{eqnarray}
where
\begin{eqnarray}
\mu_N &=& \sqrt{{K_{F_N}}^2+ {M_N^*}^2}+ 3g_\omega^q \bar\omega \label{mun}
 \\
\mu_\Lambda &=& \sqrt{{K_{F_\Lambda}}^2+ {M_\Lambda^*}^2}+
2g_\omega^q \bar\omega+ g_\phi^s \bar\phi \label{mul}
 \\
\mu_\Xi &=& \sqrt{{K_{F_\Xi}}^2+ {M_\Xi^*}^2} +g_\omega^q \bar\omega
+ 2g_\phi^s \bar\phi \label{mux}
\end{eqnarray}
Substituting Eqs.(\ref{mun})-(\ref{mux}) into Eq.(\ref{cheq}), we obtain the following
condition for the chemical equilibrium among $\Xi$s, $\Lambda$s, and
the nucleons:
\begin{eqnarray}
2 \sqrt{{K_{F_\Lambda}}^2+ {M_\Lambda^*}^2} = \sqrt{{K_{F_N}}^2+
{M_N^*}^2} + \sqrt{{K_{F_\Xi}}^2+ {M_\Xi^*}^2}. \label{cheq2}
\end{eqnarray}

One usually defines a strangeness fraction $f_s$ as
\begin{eqnarray}
f_s \equiv \frac{\rho_\Lambda + 2\rho_\Xi}{\rho_B} \label{dfs}
\end{eqnarray}

Given $\rho_B$ and $f_s$, we determine $\rho_N$, $\rho_\Lambda$,
$\rho_\Xi$ by Eqs.(\ref{drhob}), (\ref{cheq2}) and (\ref{dfs}).

\subsection {Single $\Lambda$ hypernuclei}

We now turn to hypernuclei in the IQMDD model.
$\Lambda$ hypernucleus is treated as a system of many nucleons and one $\Lambda$ hyperon
which interact through exchange of $\sigma, \omega$ mesons.
Similar to the QMC model \cite{QMC1},
one constructs a relativistic Lagrangian density at the hadronic level in the following form
\begin{eqnarray}
\mathcal{L} &=&
\bar{\psi}_N[\gamma^{\mu}(i\partial_\mu-g_\omega^N\omega_\mu-\frac{g_\rho^N}{2}\tau\cdot\rho_\mu-
\frac{e}{2}(1+\tau_3)A_\mu)-M_N^*(\sigma)]\psi_N
\nonumber\\&&+\bar{\psi}_\Lambda[\gamma^{\mu}(i\partial_\mu-g^\Lambda_\omega\omega_\mu
)-M_{\Lambda}^*(\sigma)]\psi_\Lambda \nonumber\\&&
+\frac{1}{2}\partial^\mu\sigma\partial_\mu\sigma-U(\sigma)
-\frac{1}{4}V_{\mu\nu}V^{\mu\nu}+\frac{1}{2}m_{\omega}^2\omega_\mu\omega^\mu \nonumber\\
&&-\frac{1}{4}b^{\mu\nu}b_{\mu\nu}+\frac{1}{2}m^2_\rho\rho^\mu\rho_\mu-
\frac{1}{4}F^{\mu\nu}F_{\mu\nu}
\end{eqnarray}
where the $\psi_N$ and $\psi_\Lambda$ are the Dirac spinors for the
nucleon and the $\Lambda$ hyperon, the strength tensors of the vector mesons and electromagnetic field are defined as:
$V_{\mu\nu}=\partial_\mu\omega_\nu-\partial_\nu\omega_\mu$,
$b_{\mu\nu}=\partial_\mu\rho_\nu-\partial_\nu\rho_\mu$,
$F_{\mu\nu}=\partial_\mu A_\nu-\partial_\nu A_\mu$.
$g_\omega^N$ and $g_\rho^N$ are the coupling constants between
nucleon and $\omega$ meson, and nucleon and $\rho$ meson respectively. They satisfy $g_\omega^N = 3g^q_\omega$ and $g_\rho^N
= g^q_\rho$ \cite{Wu5}.
Here the hypernucleus system is essentially composed of nucleons and there is only one $\Lambda$ hyperon,
obviously the it is no longer in any chemical equilibrium.
Therefore we take into account the contributions from $\rho$ mesons in nuclear interactions
and ignore those from strange mesons.
Under MFA with assumed spherical symmetry, the Lagrangian can be further simplified as
\begin{eqnarray}
\mathcal{L}_{RMF} =
&\bar{\psi}_N&[i\gamma^{\mu}\partial_\mu-M^*_N(\bar\sigma)-\big{(}g_\omega^N
\bar\omega(r) + g_\rho^N \frac{\tau_3}{2} \bar\rho(r) +
\frac{e}{2}(1+\tau_3)A_0(r) \big{)}\gamma_0 ]\psi_N \nonumber \\
&+&
\bar{\psi}_\Lambda[i\gamma^{\mu}\partial_\mu-M^*_\Lambda(\bar\sigma)-g_\omega^{\Lambda}
\bar\omega(r)\gamma^0]\psi_\Lambda \nonumber \\
&-& \frac{1}{2}(\nabla \bar\sigma(r))^2 \hskip 0.0in -
U(\bar\sigma) + \frac{1}{2} [(\nabla \bar\omega(r))^2 + m_\omega^2
\bar\omega(r)^2 ]
 \nonumber
\\&+& \frac{1}{2}[(\nabla \bar\rho(r))^2 + m_\rho^2 \bar\rho(r)^2 ] +
\frac{1}{2}(\nabla A_0(r))^2 \label{lgnucleusmft}
\end{eqnarray}
where $A_0$ denotes the electric field.
From the Lagrangian density given by Eq.(\ref{lgnucleusmft}), using the Euler-Lagrange equation we obtain the Dirac equation
for nucleon and hyperon as follows:
 \begin{eqnarray}
[-i\vec\alpha \cdot \vec\nabla + \beta M^*_N(\bar\sigma)+g_\omega^N
\bar\omega(r) + g_\rho^N \frac{\tau_3}{2} \bar\rho(r) +
\frac{e}{2}(1+\tau_3)A_0(r) ]\psi_j = \varepsilon_j  \psi_j,
\end{eqnarray}
 \begin{eqnarray}
[-i\vec\alpha \cdot \vec\nabla + \beta
M^*_\Lambda(\bar\sigma)+g_\omega^\Lambda \bar\omega(r)]\psi_\Lambda
= \varepsilon_\Lambda \psi_\Lambda.
\end{eqnarray}

The Klein-Gordon equations for the mesons and photon can be written as
 \begin{eqnarray}
(-\triangle + m_\sigma^2)\bar\sigma = - \frac{\partial
M_N^*}{\partial \bar\sigma} \rho_s - \frac{\partial
M_\Lambda^*}{\partial \bar\sigma} \rho_s^\Lambda -
b\bar\sigma^2-c\bar\sigma^3, \label{eqmftn}
\end{eqnarray}
 \begin{eqnarray}
(-\triangle + m_\omega^2)\bar\omega = g_\omega^N \rho_B +
g_\omega^\Lambda \rho_B^\Lambda,
\end{eqnarray}
 \begin{eqnarray}
(-\triangle + m_\rho^2)\bar\rho = \frac{g_\rho^N}{2} \rho_3,
\end{eqnarray}
 \begin{eqnarray}
-\triangle A_0 = e \rho_p, \label{eqmfta}
\end{eqnarray}
where $\rho_s(\rho_s^\Lambda), \rho_B(\rho_B^\Lambda) $ and $\rho_p$
are the densities of scalar, baryon and proton in the hypernucleus, respectively.
$\rho_3$ is the difference between the neutron and proton densities.
The above coupled equations (\ref{eqmftn})-(\ref{eqmfta}) can be self-consistently solved once the effective masses of hadrons are obtained from Eq.(\ref{bage}).

\section{Determination of the parameters of the model}

Before setting off for numerical calculations, let us determine the parameters in IQMDD model.
First, we fix nucleon mass $M_N=939$ MeV and choose $m_\omega=783$ MeV and $m_\sigma=509$ MeV as
that of Ref. \cite{parameters} and $m_{\sigma^*}=975$ MeV, $m_\phi=1020$ MeV\cite{Song}.
By fixing the bag constant at $B=174$ MeV fm$^{-3}$, one obtains $b=-1460$ MeV and $c=2.7$ as in Ref.\cite{Wu2}
Obviously, the properties at the saturation point must be reproduced by the model.
Symmetric nuclear matter saturates at a density $\rho_0=0.15$ fm$^{-3}$  with a binding energy per particle $E/A= -15$ MeV
at zero temperature, and the compression constant is about $K(\rho_0)=210$ MeV.
We therefore fix $ g_\omega^q=2.44$, $g^q_\sigma=4.67$ to explain the above data.
In addition, symmetry energy coefficient 33.2 MeV is used to fix $g_\rho^q=9.07$.
One also uses $m_{s0}$ = 162 MeV to fit free $\Lambda$ mass $M_\Lambda$ = 1116 MeV.
For $\Xi$, the same parameters are made use of as those for $\Lambda$ and
one subsequently obtains $M_\Xi$ = 1306 MeV, which is very close to the theoretical value $M_\Xi$ = 1318 MeV.
For simplicity, we adopt the above value in our calculations.

Now we come to the coupling constants concerning strange degree of freedom.
Since it is no coupling between the s quark and vector meson $\omega$ due to the OZI rule
\cite{OZI} and the $\sigma^*$ and $\phi$ mesons couple only to hyperons,
there are three coupling constants left: $g_\sigma^s, g_{\sigma^*}^s, g_\phi^s$.
As for the coupling constant $g_\sigma^s$, we use the experimental value
$E_\Lambda^{(N)}$  which is the energy of one single $\Lambda$ in symmetric nuclear matter at saturation.
According to the analysis of Bouyssy \cite{Bouyssy} and of Hausmann and Weise \cite{Hausmann}
we take $E_\Lambda^{(N)} = -28$ MeV. One subsequently obtains $g_\sigma^s=0.2$
Using this coupling parameter, we obtain the theoretical energy of one single $\Xi$ in symmetric nuclear matter,
$E_\Xi^{(N)}$= -13 MeV, reasonably close to the experimental value -18 MeV \cite{-18MeV}.
Since the $\sigma^*$ and $\phi$ mesons couple only to the hyperons,
it is reasonable in the quark model to set $g_{\sigma^*}^q = g_\phi^q = 0$ (q = u, d).
We fix $g_\phi^s$ by using the SU(6) relation $g_\phi^s/g_\omega^q = -\sqrt{2}$.
As for $g_{\sigma^*}^s$, we follow the estimation for $\Lambda$-$\Lambda$ interaction energy made by Schaffner \cite{Schaffner}.
Denoting the potential depth of a single nucleon in a nucleon ``bath" at saturation density $\rho_0$ by $U_N^{(N)}$,
the potential depth of a single $\Lambda$ in a ``$\Lambda$" bath at $\rho_\Lambda \simeq 0.5 \rho_0$ by $U_\Lambda^{(\Lambda)}$,
they obtained
\begin{eqnarray}
\frac{U_\Lambda^{(\Lambda)}}{U_N^{(N)}}= \frac{1}{2}
\frac{(1/4)V_{\Lambda \Lambda}}{(3/8)V_{NN}},
\end{eqnarray}
where $V_{NN}\simeq$ 6-7 MeV, $U_N^{(N)}=80$ MeV.
From the old experimental data \cite{Y-Y old 1}-\cite{Y-Y old 3},
$V_{\Lambda\Lambda} \equiv \Delta B_{\Lambda \Lambda} \simeq 4-5$ MeV,
we obtain $U^{(\Lambda)}_\Lambda \simeq 20$ MeV.
If  one takes
$V_{\Lambda\Lambda} = 1.01 $ MeV from the the new data \cite{Y-Y new},
$U^{(\Lambda)}_\Lambda \simeq 5$ MeV.
The coupling constant $g^s_{\sigma^*} = 7.12$ is obtained once we
fit the potential depth $U_\Lambda^{(\Lambda)} \simeq 20$ MeV.
The above value was estimated according to a stronger $\Lambda$-$\Lambda$ interaction,
it will be referred to hereafter as strong Y-Y interaction.
If one uses the new value $\Delta B_{\Lambda\Lambda} \simeq 1$ MeV,
$g^s_{\sigma^*}= 2.83$ is obtained instead.
It will be therefore referred to as weak Y-Y interaction below.
Song et al. had some calculations on the potential depth in Ref. \cite{Song}.
We note that in our calculation we do not take into account the
progress in the reanalysis of double Lambda events \cite{Y-Y 2010}.
Those reanalyses indicate that the potential depth of
$U^{(\Lambda)}_{\Lambda}$ in the case of weak of Y-Y interaction may be
even shallower than that 5 MeV.

\section{Numerical result of the model}

First we discuss the saturation properties of the multi-$\Lambda$
strange nuclear matter with different ratios $f_s$.
Some discussions on multi-$Lambda$ matter in
bulk can be found in Refs. \cite{Multi-lambda1}-\cite{Multi-lambda3}.
As usual, we subtract the baryon masses in the total energy per baryon of
the strange matter given by Eq.(\ref{etotal}) and study the binding energy per baryon expressed as
\begin{eqnarray}
E/B= (\epsilon_{tot}-M_N\rho_N-M_\Lambda\rho_\Lambda  - M_\Xi \rho_\Xi)/\rho_B
\end{eqnarray}
The calculated results are summarized in Figs. 1-2.
In Fig.1, we plotted the binding energy per baryon versus the baryon density for IQMDD model at different $f_s$ values.
The minimum point of each curve corresponds to the
stability point of strange hadronic matter.
To study the stability of the system more transparently, we present the minimum of $E/B$
versus strangeness fraction $f_s$ in Fig.2.
From Fig.2, one notes that E/B possesses negative minimum up to $f_s$ = 0.8.
It implies that systems containing up to $80\%$
Lambdas will still be stable against particle decay.
As $f_s$ increases, the saturation curve becomes deeper first and then goes shallower.
The lowest minimum occurs around $f_s$ = 0.1.
Compared to ordinary symmetric nuclear matter,
strange hadronic matter with $f_s = 0.1$ has an additional binding energy of $\sim 0.3$ MeV.
The increase in binding energy is comparable to
that in the modified QMC model, which is about 0.6 MeV \cite{Wang2}.

Fig.3 shows the effective baryon masses as a function of baryon density when they are in ``bath'' of bulk hadronic matter.
Curves with labels $f_s=0.0$ show results of symmetric ordinary nuclear matter,
and those with lables $f_s=1.0$ give corresponding results of nucleon-$\Lambda$-$\Xi$ mixture matter.
The solid lines stand for nucleons, the dashed lines are for $\Lambda$s,
and the dotted lines are for $\Xi$s.
In all cases, the effective baryon masses drop down
monotonously when nuclear density increases.
As expected, the effective baryon masses satisfy the following order:
$M^*_{\Xi}$ is the largest, followed by $M^*_{\Lambda}$ and $M^*_{N}$ is the smallest.
First, it is natural to understand the above mass ordering
because the coupling between $\Xi$ and $\sigma$ meson is about one third of that
between the nucleon and $\sigma$ meson.
As a result, for hadrons in ``bath'' of symmetric nuclear matter, the $\Xi$-$N$ coupling is weaker than the $N-N$ coupling.
since the contributions from strange mesons are irrelevant in this case.
One can see from Fig.1 that the effective
$\Lambda$ mass in the $f_s=1.0$ case is almost same as that in the case of $f_s=0.0$.
On the other hand, when strangeness fraction
increases the effective nucleon mass increases significantly,
while the effective $\Xi$ mass decreases.
This is a result of competition between the following two factors.
In the first place, when the strangeness fraction increases, the number of non-strange quarks becomes smaller,
which supresses any hadron-hadron interaction mediated by the non-strange mesons.
Secondly, the number of s quark increases with strangeness fraction,
and subsequently amplifies the hadron-hadron interaction mediated by the strange mesons.
In case of nucleons, since it does not contain any strange quark, it is hardly affected by the second factor.
Therefore when the strangeness fraction increases, interaction between nucleons is suppressed, results in decreasement of its effective mass.
On the contrary, since $\Xi$ contains two strange quarks, its effective mass is essentially determined by the first factor.
$\Lambda$ hyperon stays in the middle, the two effects are more or less balanced with one another
in such a way that its effective mass is almost not affected by strangeness fraction.

The calculated results of saturation properties of nucleon-$\Lambda$-$\Xi$ mixture matter are summarized in Figs. 4-7.
Such systems consist of symmetric nuclear matter in equilibrim with $\Lambda$s and $\Xi$s.
Due to the condition of chemical equilibrium, namely Eq.(\ref{cheq2}),
the particle density of $\Lambda$s and that of $\Xi$s are not at all independent.
Given the values of $\rho_B$ and $f_s$, particle densities of $\Lambda$s and $\Xi$s are fully fixed by Eqs.(\ref{cheq2}) and (\ref{dfs}).

In Fig.4, we show the binding energy per baryon $E/B$ versus baryon
density $\rho_B$ at various strangeness fractions $f_s$ calculated with weak Y-Y interaction.
It is seen that the saturation curve gets
shallower and shallower with increasing strangeness fraction $f_s$.
And there is no negative minimum in the saturation curve when $f_s$
value is larger than about 1.2. The results indicate that the
strange hadronic matter with the weak Y-Y interaction is less stable
than the normal nuclear matter and becomes unstable when the $f_s$ is over 1.2.

We should emphasize that numerical calculations above are based on the experimental data of the weak Y-Y interaction.
If we would like to investigate the influence of the strength of Y-Y interaction on
the strange hadronic matter,  it is necessary to investigate strange
hadronic matter both with the strong Y-Y interaction and the weak Y-Y interaction.
In Fig.5, the energy per baryon vs baryon density in the strange hadronic matter
with various of $f_s$ calculated with a strong Y-Y interaction is shown.
One can see that the situation in this case is very different from that of the weak Y-Y interaction case as shown in Fig.4.
In general, strange hadronic matter is more stable than ordinary nuclear matter with strong Y-Y interaction.

To study the stability of the systems,
we again minimize the $E/B$ with respect to $\rho_B$ at each strangeness fraction $f_s$,
and we present minimum of $E/B$ as a function of $f_s$ in Fig.6,
and the corresponding $\rho_B$ are plotted in Fig.7.
It is found that the minimum of energy per baryon
calculated with the strong Y-Y interaction occurs at the point $(E/B, f_s) \simeq$ (-21.5 MeV, 1.5).
Compared to the symmetric nuclear matter,
the system gets an additional binding energy per baryon of about 6.5 MeV.
This is caused by the strong attraction between the $\Xi$s.
The corresponding saturation density in two cases are  shown in Fig.7.
Again the difference between the strong
and the weak Y-Y interactions manifested itself.
From Figs.6-7, for the strong Y-Y interaction we can find that the most deeply
bound state appears at baryon density $\rho_B \simeq 0.40$ fm$^{-3}$
and with strangeness fraction $f_s \simeq 1.5$ where $\Xi$ dominates.
However, if the $\Lambda$-$\Lambda$ interaction is weak,
the strange hadronic matter with any strangeness fraction
is even less stable than normal nuclear matter.
The larger the strangeness fraction is, the less stable the system is.
The minimized energy for given $f_s$ increases with increasing $f_s$.
There is no negative minimum when $f_s$ is larger than about 1.2.

After showing the numerical results for strange hadronic matter,
it is interesting to perform the calculations self-consistently for $\Lambda$ hypernuclei in the IQMDD model
without further adjusting the parameters.
In Fig.8, we show the effective masses of
the nucleon and $\Lambda$  as well as the baryon densities
calculated for (a) $^{17}_\Lambda $O, (b) $^{41}_\Lambda $Ca and (c) $^{209}_\Lambda $Pb.
The results are for the $1s_{1/2}\,\, \Lambda$ state,
where effects of the $\Lambda$ hyperon on the whole system are expected to be the largest.
The effective masses in the all three hypernuclei ($^{17}_\Lambda $O, $^{41}_\Lambda $Ca, $^{209}_\Lambda
$Pb) behave in a similar manner as the distance $r$ from the center of each nucleus increases (the baryon density decreases).
The calculated $\Lambda$ single-particle energies for the closed-shell core plus
one $\Lambda$ configuration are listed in Table. 1.
One can easily see that spin-orbit splittings in the present model are very small for all hypernuclei.
Its magnetude tends to be even smaller as the baryon
density increases or the atomic number increases.
In Fig.9, we show single-particle energies of $\Lambda$ hyperon for these hypernuclei in the IQMDD model.
In order to reduce finite size effects such as surface effect, the calculated energies are presented as a function of $A^{-2/3}$.
For $A\rightarrow \infty$,  the $1s_{1/2}$ $\Lambda$ energies converge asymptotically to 25 MeV,
which is close to the binding energy of a single $\Lambda$ in infinite matter 28 MeV.
\newline

\begin{tabular}
{p{1.2cm}p{1.2cm}p{2.2cm}p{1.2cm}p{2.2cm}p{1.2cm}p{1.2cm}p{2.2cm}p{1.4cm}p{1.8cm}}
\multicolumn{10}{c}{TABLE 1. The $\Lambda$ single-particle energies
(in MeV) for $^{17}_\Lambda$O, $^{41}_\Lambda$Ca and $^{209}_\Lambda$Pb.} \\
\hline\hline
 & $^{17}_{\Lambda}$O & $^{16}_{\Lambda}$O(Exp) &$^{41}_{\Lambda}$Ca &$^{40}_{\Lambda}$Ca(Exp)  & $^{49}_{\Lambda}$Ca &
 $^{91}_{\Lambda}$Zr  & $^{89}_{\Lambda}$Y(Exp) &$^{209}_{\Lambda}$Pb &$^{208}_{\Lambda}$Pb(Exp)
\\ \hline

$1s_{1/2}$ &-12.4  &-12.5   &-15.8       &-18.7      &-16.9  &-19.3   &-23.1  &-21.4   &-26.3   \\
$1p_{3/2}$ &-4.3  &         &-12.5       &           &-13.3  &-16.9  &         &-18.6    &   \\
$1p_{1/2}$ &-3.7  &-1.8(1p)   &-12.1      &-12.0(1p)  &-13.0  &-16.8   &-16.5(1p)  &-18.6 & -21.9(1p)  \\

$1d_{5/2}$ &     &           &-8.2       &           &-8.5    &-12.6   &         &-15.2   &  \\
$2s_{1/2}$ &  &              &-7.5       &           &-7.8    &-10.8   &         &-14.8  &  \\

$1d_{3/2}$ &  &   &-8.0       &  &-8.3  &-12.5   &-9.1(1d)  &-15.2  &-16.8(1d)   \\
$1f_{7/2}$ &  &  &           &  &  &-7.4   &  &-11.0  &   \\
$2p_{3/2}$ &  &  &           &  &  &-5.8   &  &-10.8  &    \\
$1f_{5/2}$ &  &   &          &  &  &-7.3   &-2.3(1f)  &-10.9  &-11.7(1f)  \\
$2p_{1/2}$ &  &  &           &  &  &-5.7   &  &-10.8  &  \\

$1g_{9/2}$ &  &   &          &  &  &   &  &-6.7  &   \\
$1g_{7/2}$ &  &  &           &  &  &   &  &-6.6  &-6.6(1g)   \\



\hline\hline
\end{tabular}

\begin{tabular}
{p{1.7cm}p{2.0cm}p{2.0cm}p{2.0cm}p{2.0cm}p{2.0cm}p{2.0cm}}
\multicolumn{7}{c}{TABLE 2. Binding energy per baryon, $-E/A$ (in
MeV), rms charge radius ($r_{c}$),  } \\
\multicolumn{7}{c}{ and rms radii of the $\Lambda$ ($r_\Lambda$),
neutron ($r_n$) and proton ($r_p$).
} \\
\hline\hline

 &$\Lambda$ state &-E/A  &$r_{c}$  &$r_\Lambda$ &$r_n$  &$r_p$
\\\hline

$^{17}_{\Lambda}$O  &$1s_{1/2}$ &8.83  &2.63  &2.56  & 2.48  &2.51 \\
$^{17}_{\Lambda}$O  &$1p_{3/2}$ &8.25  &2.64  &3.35  & 2.49  &2.52\\
$^{16}$O            &           &8.10  &2.61  &      &2.45   &2.48 \\
\hline

$^{41}_{\Lambda}$Ca  &$1s_{1/2}$ &8.94  &3.33  &3.05  & 3.21  &3.24 \\
$^{41}_{\Lambda}$Ca  &$1p_{3/2}$ &8.57  &3.34  &3.61  & 3.21  &3.25 \\
$^{40}$Ca            &           &8.35  &3.42  &      &3.29   &3.33 \\
\hline

$^{49}_{\Lambda}$Ca  &$1s_{1/2}$ &9.11  &3.42  &3.11  & 3.51  &3.33 \\
$^{49}_{\Lambda}$Ca  &$1p_{3/2}$ &8.96  &3.42  &3.69  & 3.51  &3.33 \\
$^{48}$Ca            &           &8.71  &3.49  &      &3.62   &3.40 \\
\hline

$^{91}_{\Lambda}$Zr  &$1s_{1/2}$ &8.79  &4.19  &3.64  & 4.34  &4.11 \\
$^{91}_{\Lambda}$Zr  &$1p_{3/2}$ &8.72  &4.20  &4.21  & 4.35  &4.12\\
$^{90}$Zr            &           &8.58  &4.23  &      &4.38   &4.15 \\
\hline

$^{209}_{\Lambda}$Pb  &$1s_{1/2}$ &7.73  &5.52  &4.07  & 5.74  &5.46 \\
$^{209}_{\Lambda}$Pb  &$1p_{3/2}$ &7.71  &5.51  &4.48  & 5.74  &5.45\\
$^{208}$Pb            &           &7.67  &5.52  &      &5.74   &5.46 \\

 \hline\hline
\end{tabular}
\newline

In Table. 2, we enumerate the calculated binding energy per baryon $-E/A$,
the RMS charge radii $r_c$, the RMS radii of the $\Lambda$,
and that of neutron and proton ($r_n$, $r_p$) respectively
for the $1s_{1/2}$ and $1p_{3/2}$ $\Lambda$ configurations.
The RMS charge radii are calculated by convolution with a proton form factor \cite{Wu6}.
For comparison, we also give these quantities for normal finite nuclei.
The differences in values for finite nuclei and hypernuclei listed in Table. 2 reflect
the effects the $\Lambda$ through the self-consistent procedure.
One can easily see that the effects of the $\Lambda$ become weaker as the atomic number becomes larger.
Regarding the effects of the $\Lambda$ on the core nucleons, we also show in Fig.10 the comparisons of nucleon
single particle energies between $^{41}_\Lambda $Ca and $^{40}$Ca for $1s_{1/2}$ $\Lambda$ state.
The existence of the $\Lambda$ causes the scalar and baryon densities to be larger,
and the scalar and vector potentials to become stronger.
As a consequence, the binding energy of nucleons in $^{41}_\Lambda$Ca are more deeper than those of $^{40}$Ca.
In addition, we also show that the scalar and vector potential strength for
$^{17}_\Lambda $O, $^{41}_\Lambda $Ca, $^{209}_\Lambda $Pb in Fig.11.

\section{Summary and discussions}
In summary, we have extended the IQMDD model to include s quark degree of freedom
and use the model to discuss the properties of strange hadronic matter and those of $\Lambda$ hypernuclei.
An uniform parameterization has been used to conduct the calculations consistently.
The properties of multi-hyperon nuclear matter, such as the density dependence of the effective baryon masses
and the stabilities of the strange hadronic matter are discussed.
From the above discussions, we arrived to the conclusion that the different Y-Y interactions result in very different systems.
It is found that the strange hadronic matter with the weak Y-Y interaction is rather loosely
bound comparing to ordinary nuclear matter.
The model is then applied to calculate physical quantities pertaining to $\Lambda$ hypernuclei,
such as binding energy per baryon, charge radii etc.
Moreover, the spin-orbit coupling for $\Lambda$ hypernuclei was found to be small as consistent with experimental observations.
The calculated results show that the IQMDD model gives reasonable description
for the properties of strange hadronic matter, as well as those of single $\Lambda$ hypernuclei.

\begin{acknowledgments}
We acknowledge funding from the National Natural Science Foundation of China (NNSFC Grants 11105072, 10979074, 10875160, 10805067 and 11035009),
and from Brazilian Foundations Funda\c{c}\~ao de Amparo \'a Pesquisa do Estado de Minas Gerais (FAPEMIG)
and Conselho Nacional de Desenvolvimento Cientit\'{\i}fico e Tecnol\'ogico (CNPq).
\end{acknowledgments}

\begin{figure}[tbp]
\includegraphics[width=14cm,height=20cm]{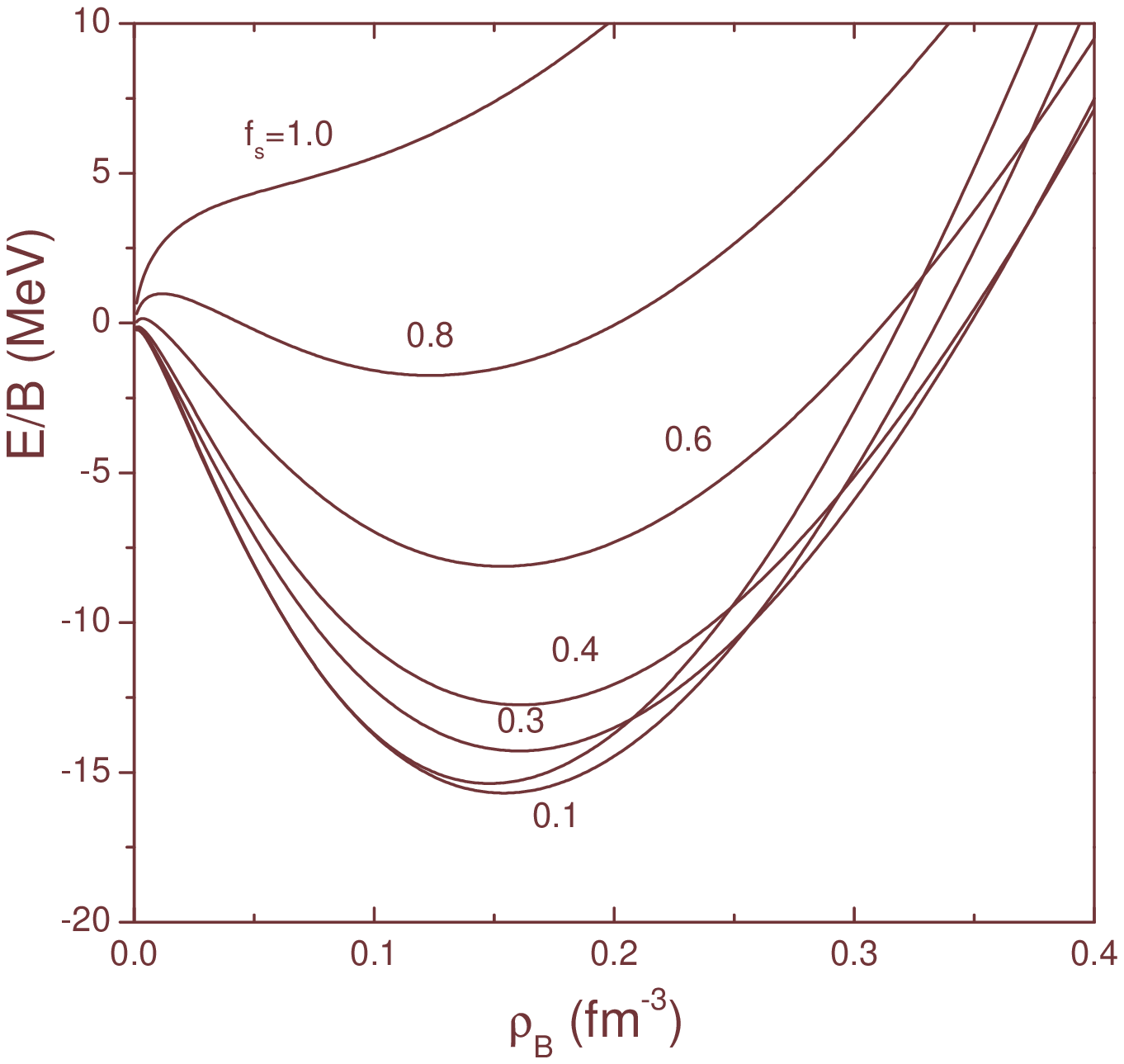}
\caption{Energy per baryon versus baryon density density in the
nucleon-Lambda mixture with various values of $f_s$ ranged from 0.0
to 1.0.}
\end{figure}

\begin{figure}[tbp]
\includegraphics[width=14cm,height=20cm]{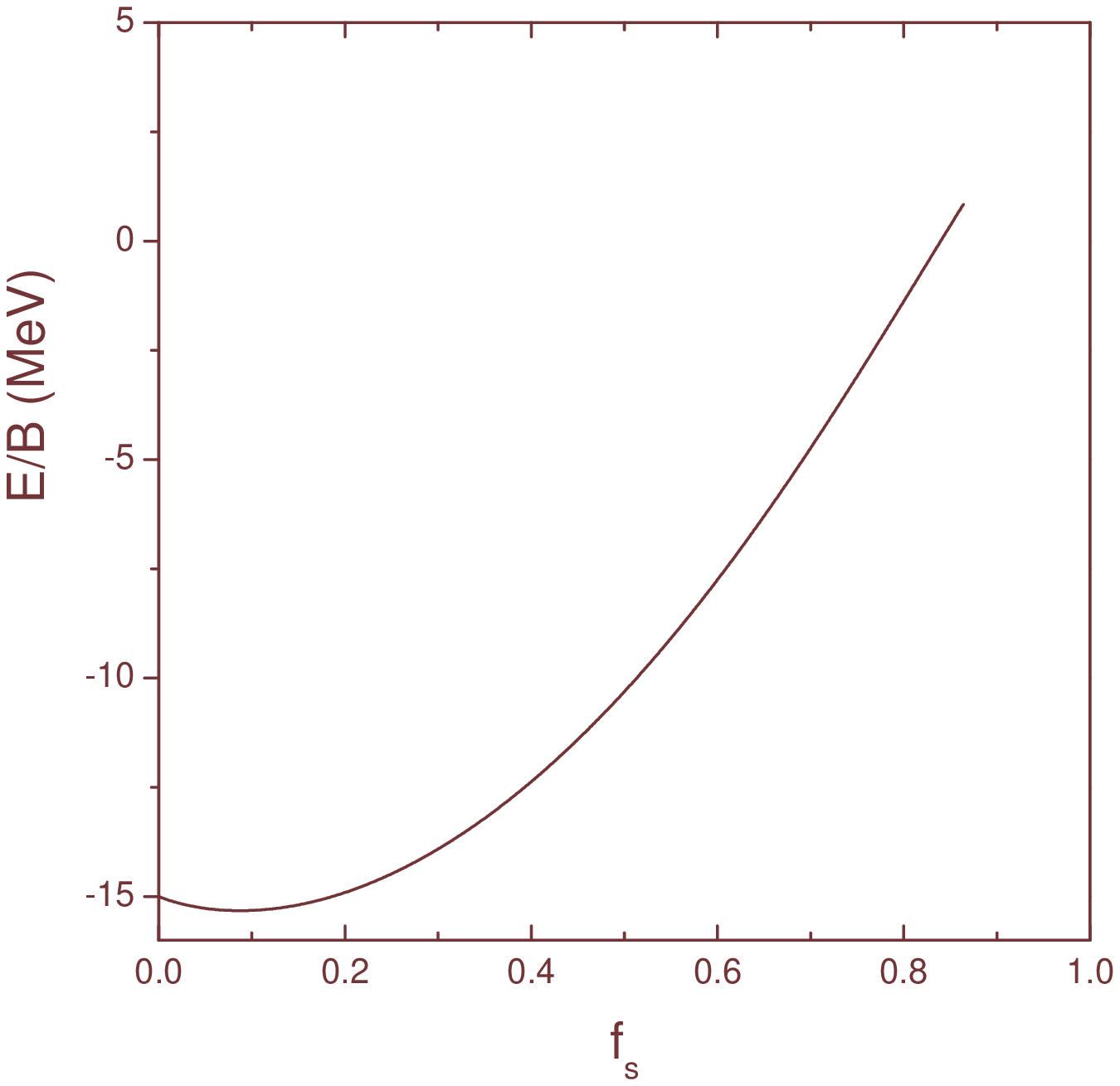}
\caption{The minimum energy per baryon in the nucleon-Lambda mixture
as a function of strangeness fraction $f_s$.}
\end{figure}

\begin{figure}[tbp]
\includegraphics[width=14cm,height=20cm]{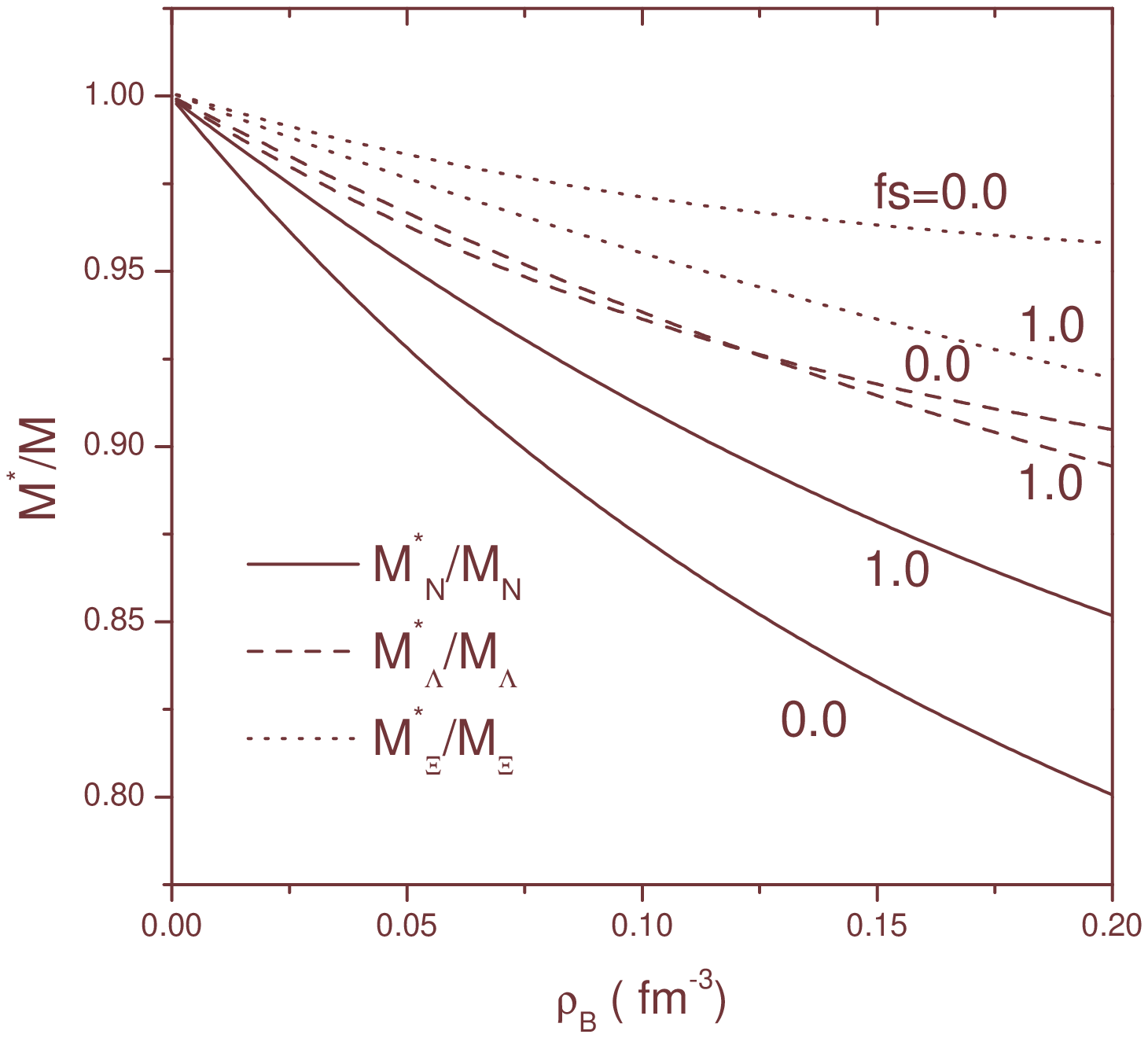}
\caption{Effective baryon masses versus baryon density in the
nucleon-$\Lambda$-$\Xi$ mixture with various $f_s$ in IQMDD model.
The solid lines stand for nucleons,  the dashed lines for lambdas
and the dotted lines for $\Xi$s. }
\end{figure}

\begin{figure}[tbp]
\includegraphics[width=14cm,height=20cm]{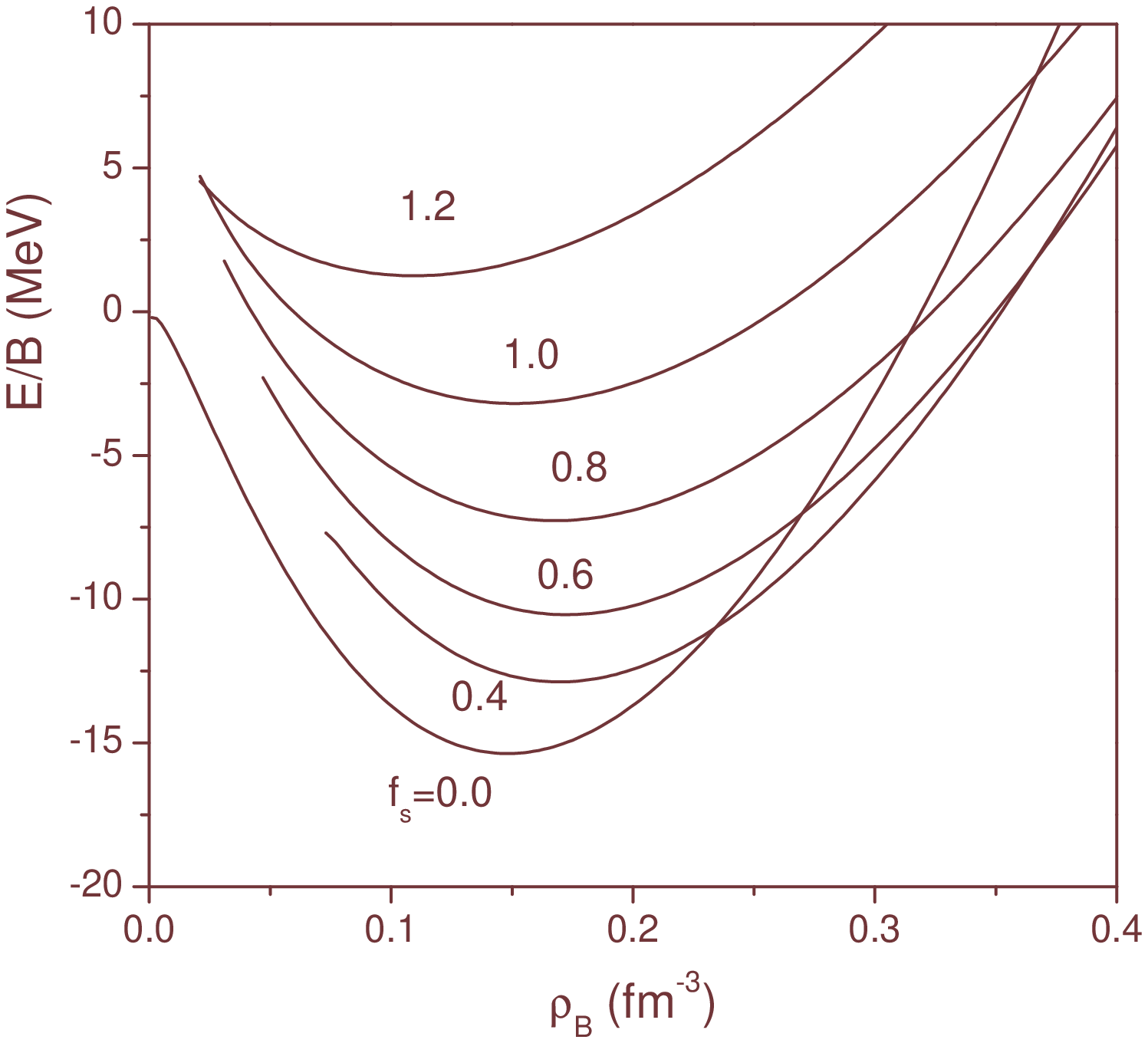}
\caption{Energy per baryon versus baryon density in the
nucleon-$\Lambda$-$\Xi$ mixture with various values of $f_s$ for the
weak Y-Y interaction. }
\end{figure}

\begin{figure}[tbp]
\includegraphics[width=14cm,height=20cm]{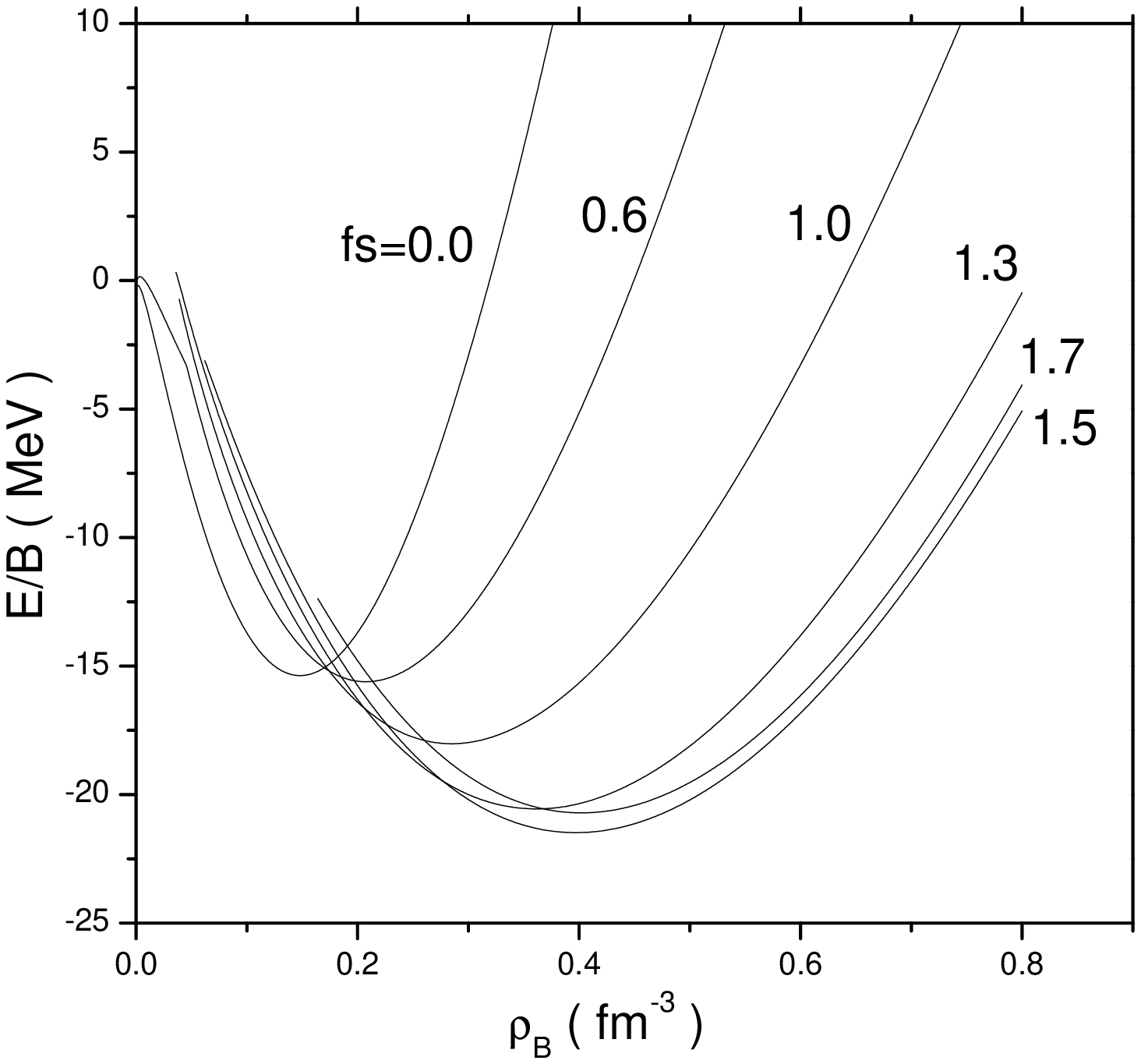}
\caption{Energy per baryon versus baryon density in the strange
hadronic matter with various values of $f_s$, calculated with the
strong Y-Y interaction. }
\end{figure}

\begin{figure}[tbp]
\includegraphics[width=14cm,height=20cm]{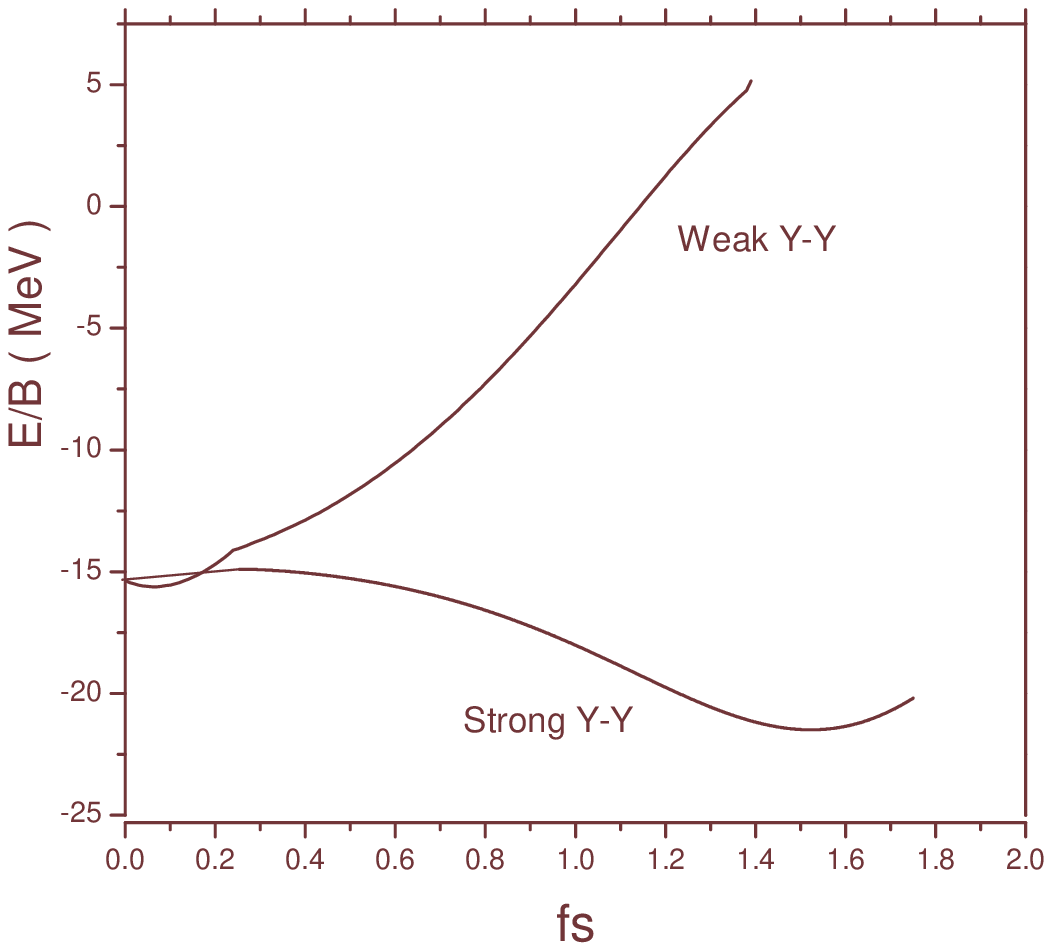}
\caption{The minimized energy per baryon (the binding energy per
baryon at saturation) in the strange hadronic matter with the weak
Y-Y interaction and the strong Y-Y interaction , as a function of
strangeness fraction $f_s$. }
\end{figure}

\begin{figure}[tbp]
\includegraphics[width=14cm,height=20cm]{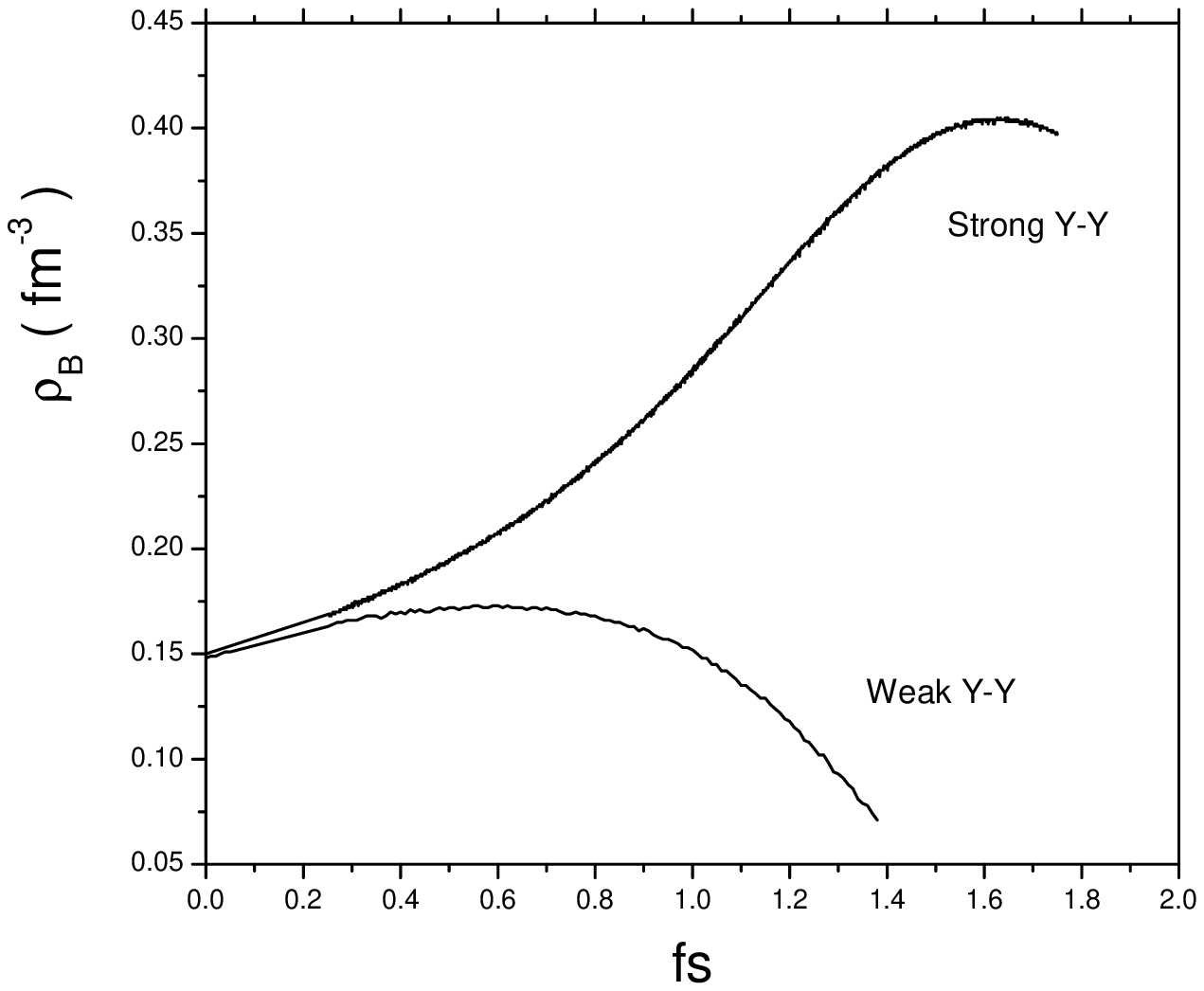}
\caption{The baryon density corresponding to the minimum energy in the strange hadronic matter with the
 weak Y-Y interaction  and the  strong Y-Y
interaction, as a function of strangeness fraction $f_s$.  }
\end{figure}

\begin{figure}[tbp]
\includegraphics[width=14cm,height=20cm]{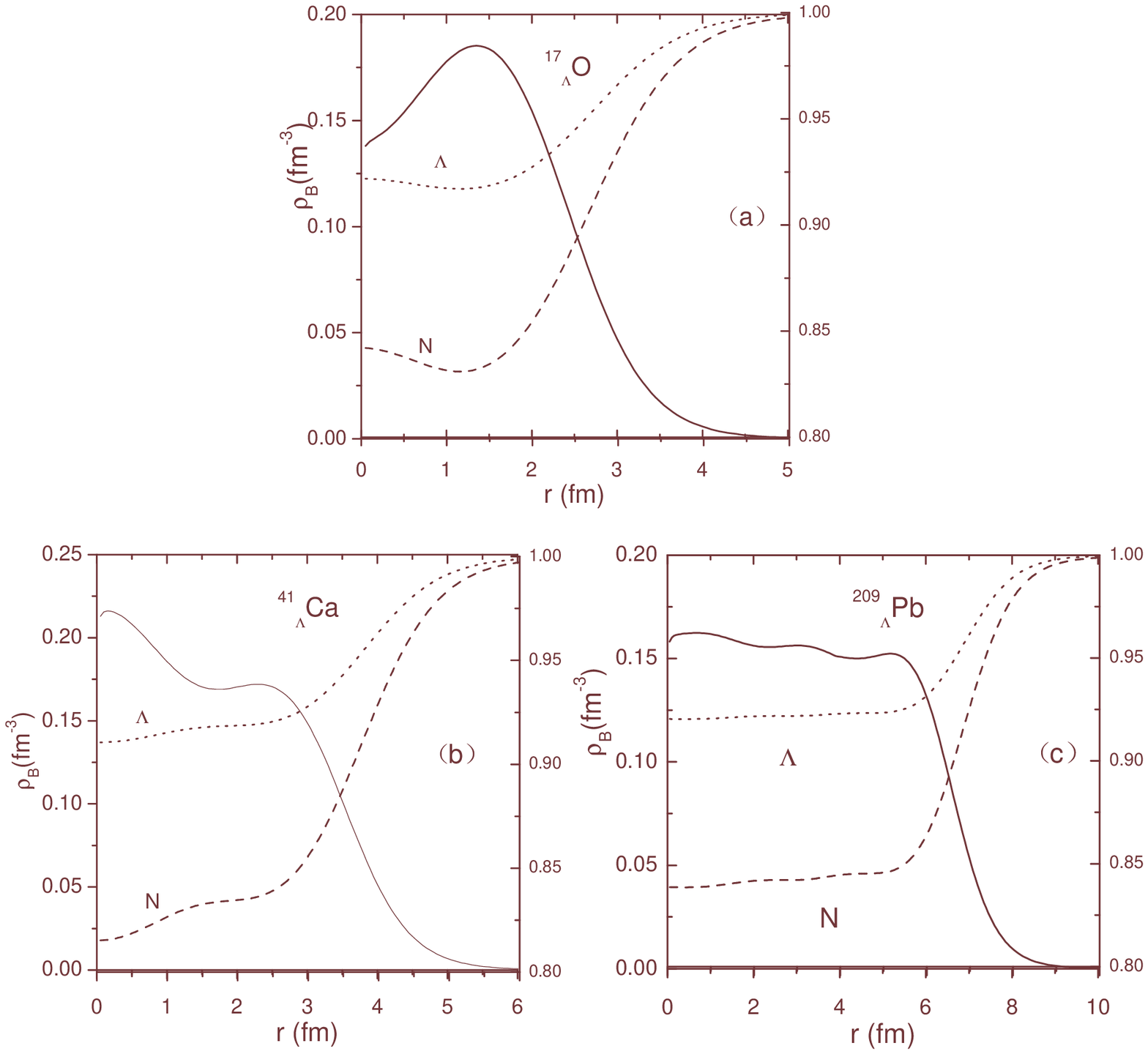}
\caption{Calculated baryon densities, $\rho_B$, and effective masses
of the nucleon and the $\Lambda$ hyperon in hypernuclei for (a):
$^{17}_\Lambda$O, (b): $^{41}_\Lambda$Ca and (c):
$^{209}_\Lambda$Pb. All cases are for the $1s_{1/2}\, \Lambda$
state.}
\end{figure}

\begin{figure}[tbp]
\includegraphics[width=14cm,height=20cm]{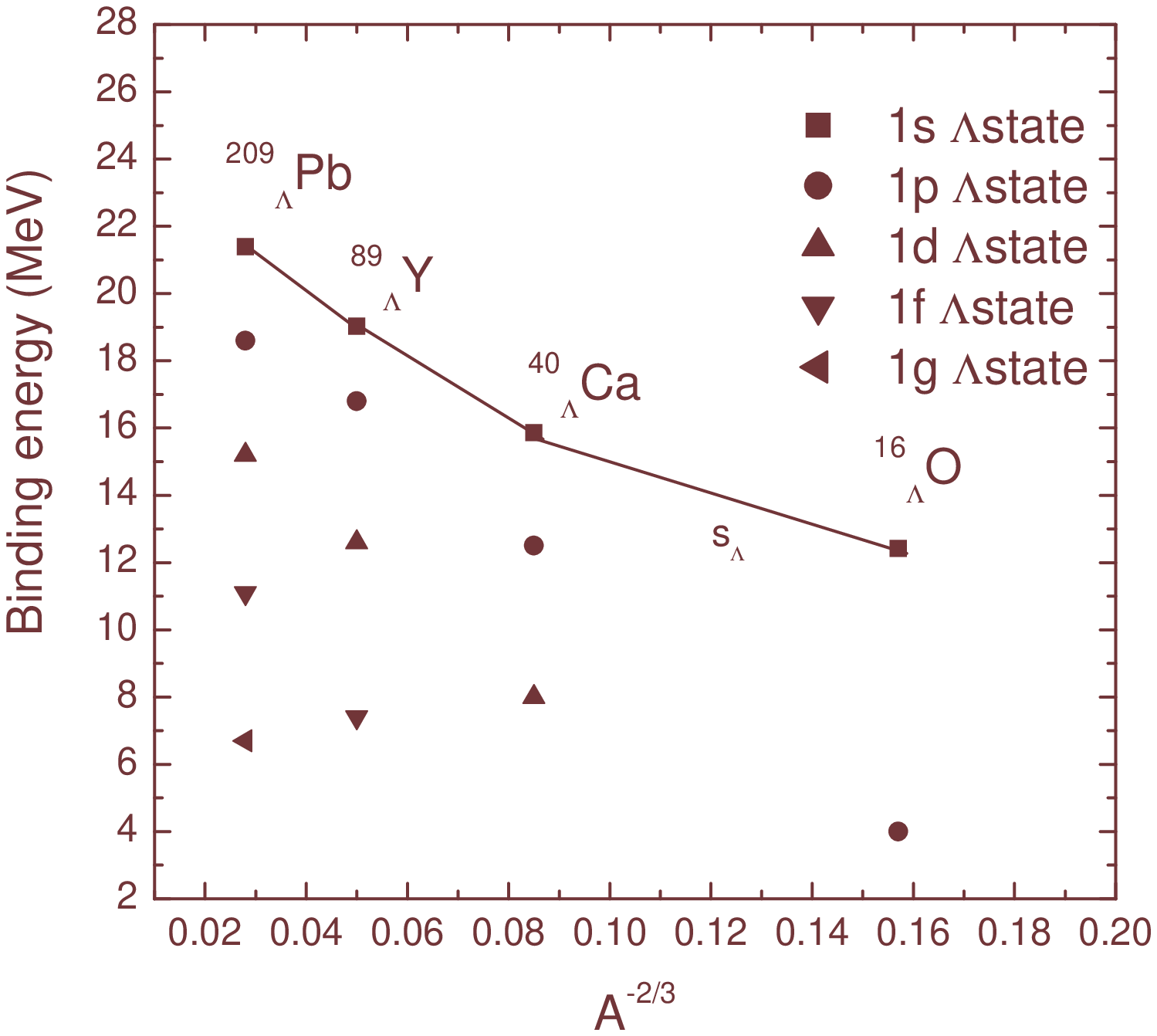}
\caption{ Single $\Lambda$ particle energies for several hypernuclei
in the IQMDD model. The energy of one single Lambda in symmetric
nuclear matter at saturation is -28 MeV. }
\end{figure}

\begin{figure}[tbp]
\includegraphics[width=14cm,height=20cm]{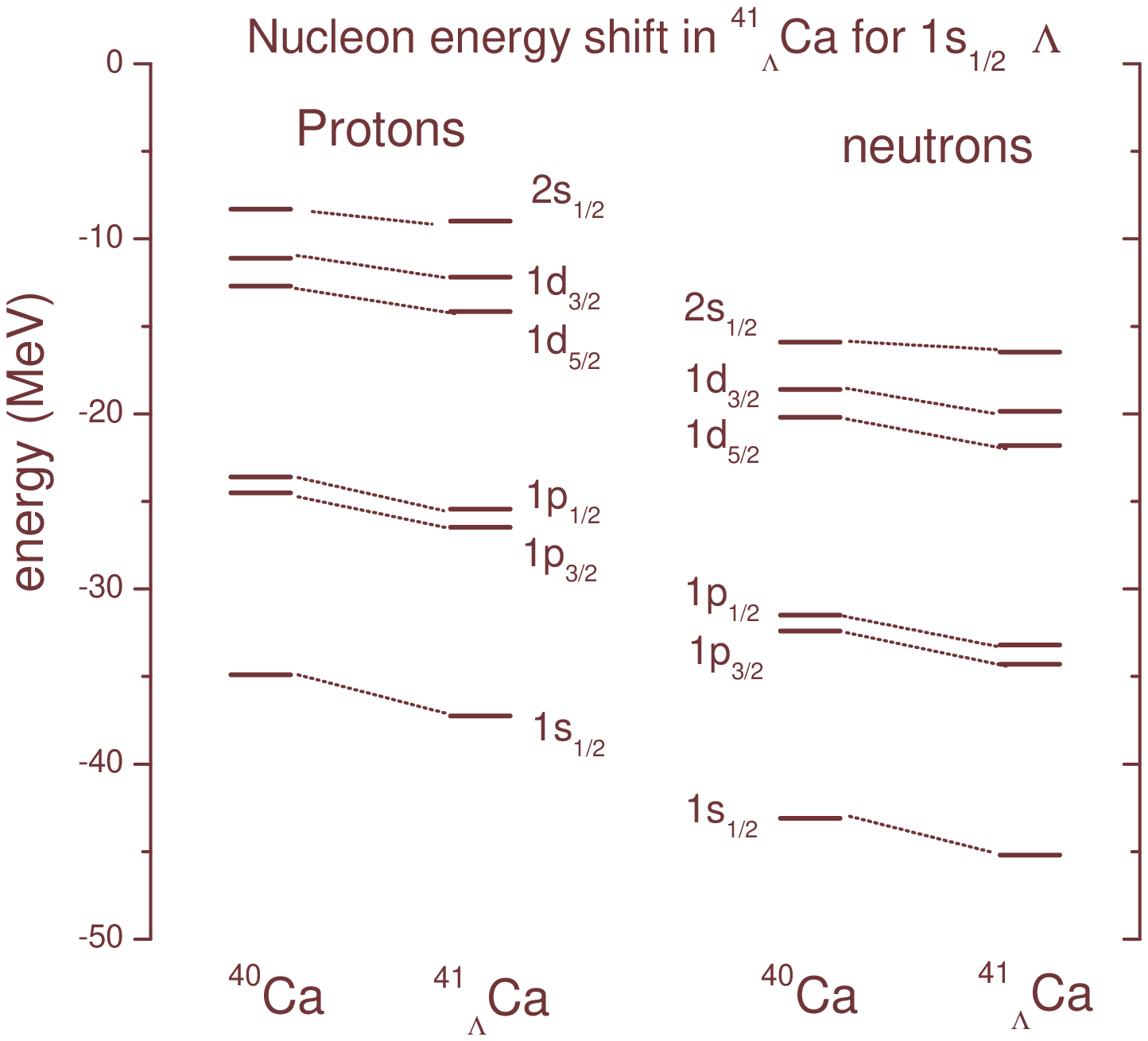}
\caption{Nucleon single particle energies for $^{40}Ca$ and
$^{41}_\Lambda$Ca for the $1s_{1/2}\, \Lambda$ state. }
\end{figure}

\begin{figure}[tbp]
\includegraphics[width=14cm,height=20cm]{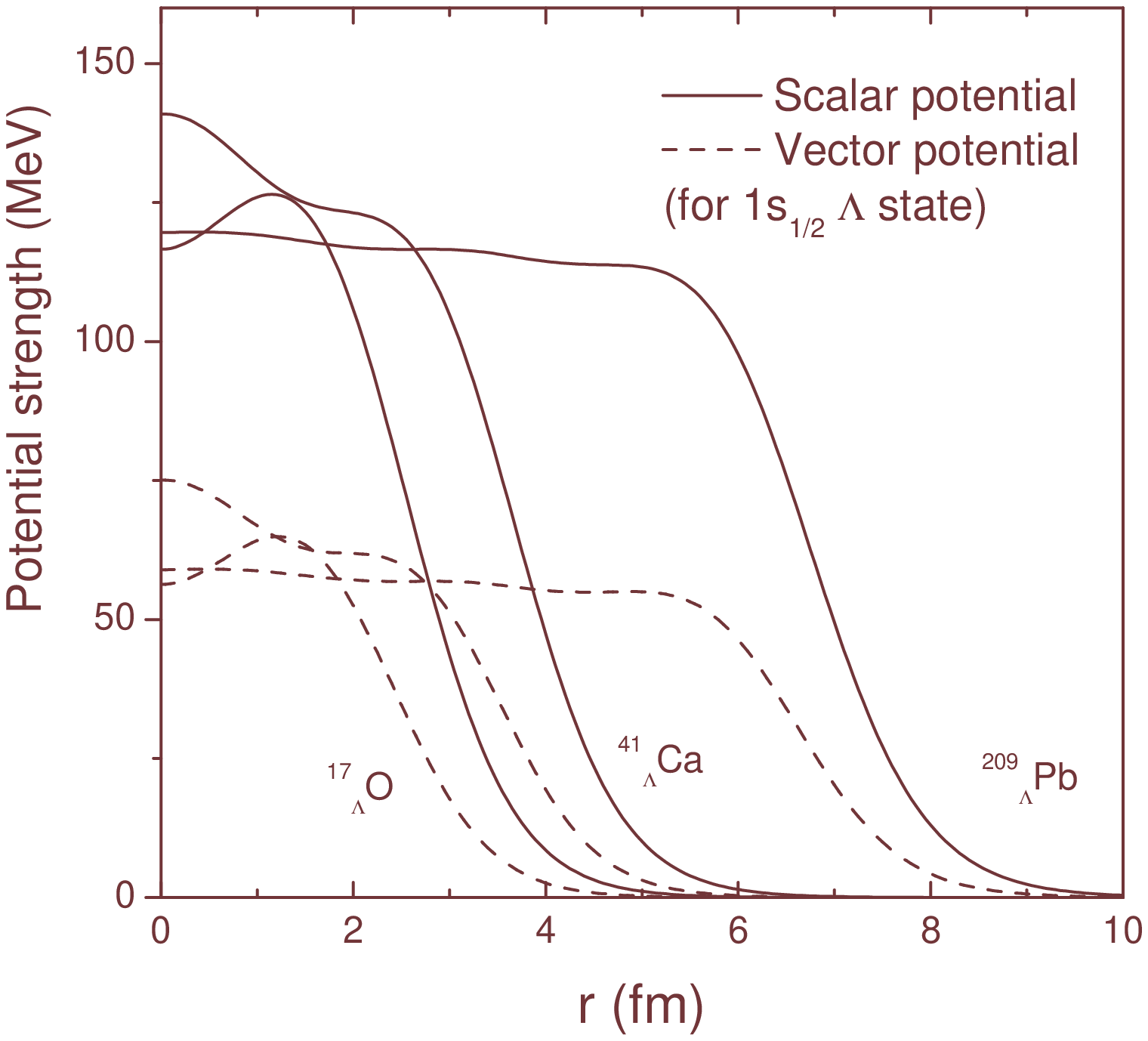}
\caption{Calculated scalar and vector potential strength for
$^{17}_\Lambda$O,  $^{41}_\Lambda$Ca and $^{209}_\Lambda$Pb. }
\end{figure}


\begin{references}

\bibitem{ns1}J. Schaffner-Bielich, H. St$\ddot{o}$ker and W. Greiner, Phys. Rev. Lett. \textbf{89} 171101 (2002).
\bibitem{ns2}Y. Yamamoto, S. Nishizaki and T. Takatsuka, Nucl. Phys. A \textbf{691}, 432 (2001).
\bibitem{ns3}P.K. Sahu and A. Ohnishi, Nucl. Phys. A \textbf{691}, 439 (2001).

\bibitem{science} The STAR Collaboration, Science \textbf{328}, 58 (2010).

\bibitem{Schaffner1} J. Schaffner and A. Gal,  Phys. Rev. C \textbf{62}, 034311 (2000).

\bibitem{Schaffner2} J. Schaffner et al.,  Phys. Rev. C \textbf{46}, 322 (1992).

\bibitem{Glendenning} N. Glendenning,  Phys. Rev. C \textbf{23}, 2757 (1981).


\bibitem{Friedman}  E. Friedman, A. Gal, Phys. Rept.  \textbf{452}, 89 (2007).


\bibitem{Wang1} P. Wang, R.K. Su, H.Q. Song and L.L. Zhang, Nucl. Phys. A \textbf{653}, 166 (1999).

\bibitem{Song} H.Q. Song, R.K. Su, D.H. Lu and W.L. Qian,  Phys. Rev. C \textbf{68}, 055201 (2003).

\bibitem{S.Zhang} S. Zhang, J. H. Chen, H. Crawford, D. Keane, Y. G. Ma, Z. B. Xu, Phys. Lett. B \textbf{684},224 (2010).

\bibitem{strangessold1}K. Ikeda, H. Bando, and T. Motoba, Prog. Theor. Phys. Suppl. \textbf{81}, 147 (1985).
\bibitem{strangessold2} M. Barranco, R. J. Lombard, S. Marcos, and S. A. Moszkowski, Physica C \textbf{44}, 178 (1991).
\bibitem{strangessold3} L. L. Zhang, H. Q. Song, and R. K. Su, J. Phys. G \textbf{23}, 557 (1997).

\bibitem{strangessold4} H.-J. Schulze, M. Baldo, U. Lombardo, J. Cugnon, and A. Lejeune, Phys. Rev. C \textbf{57}, 704 (1998).
\bibitem{strangessold5} I. Vidana, A. Polls, A. Ramos, M. Hjorth-Jensen, and V. G. J. Stoks, Phys. Rev. C \textbf{61}, 025802 (2000).

\bibitem{Hypernuclei-Exp3} H. Hotchi, Phys. Rev. C \textbf{64}, 044302 (2001).

\bibitem{Keil} C. M. Keil, F. Hofmann and H. Lenske, Phys. Rev. C \textbf{61}, 64309 (2000).


\bibitem{su3}P. Papazoglou, S. Schramm, J. Schaffner-Bielich, H. St$\ddot{o}$ker and W. Greiner, Phys. Rev. C \textbf{57}, 2576 (1998);
Ch. Beckmann, P. Papazoglou, D. Zschiesche,  S. Schramm,  H. St$\ddot{o}$ker and W. Greiner, Phys. Rev. C \textbf{59}, 411 (2001).

\bibitem{Wang2}P. Wang, H. Guo, Z. Y. Zhang, Y. W. Yu, R. K. Su, and H. Q. Song, Nucl. Phys. A \textbf{705}, 455 (2002);
W.L. Qian, R.K. Su and H.Q. Song, J. Phys. G \textbf{30} 1893, (2004).

\bibitem{QMC-hypernuclei} K. Tsushima, K. Saito and A. W. Thomas, Phys. Lett. B \textbf{411}, 9 (1997);
K. Tsushima, K. Saito, J. Haidenbauer, and A. W. Thomas, Nucl. Phys. A \textbf{630}, 691 (1998).

\bibitem{QMC-hypernuclei2} P.A.M. Guichon, A. W. Thomas and K. Tsushima, Nucl. Phys. A \textbf{814}, 66 (2008).

\bibitem{Shen2}A. Gal, Prog. Theor. Phys. Suppl. \textbf{156}, 1 (2004);
H. Shen and H. Toki,  Nucl. Phys. A \textbf{707}, 469 (2002);
H. Shen, F. Yang and H. Toki, Prog. Theor. Phys. \textbf{115}, 325 (2006).

\bibitem{Deformation} M. T. Win and K. Hagino, Phys. Rev. C\textbf{78}, 054311 (2008);
K. Hagino, M. T. Win and Y. Nakagawa (arXiv: 0903.3093).

\bibitem{Wu1} C. Wu, W. L. Qian and R. K. Su, Phys. Rev. C \textbf{77}, 015203 (2008).
\bibitem{Wu2} C. Wu, W. L. Qian and R. K. Su, Phys. Rev. C \textbf{72}, 035205 (2005);
W. L. Qian and R. K. Su, Int. J. Mod. Phys. A \textbf{20}, 1931 (2005).
\bibitem{Wu3} H. Mao, R. K. Su and W. Q. Zhao, Phys. Rev. C \textbf{74}, 055204 (2006).
\bibitem{Wu4} C. Wu  and R. K. Su, J. Phys. G \textbf{35}, 125001, (2008).
\bibitem{Wu5} C. Wu and R. K. Su, J. Phys.  G \textbf{36}, 095101, (2009).
\bibitem{Wu6} C. Wu and Z. Ren,  J. Phys.  G \textbf{37}, 105110, (2010).

\bibitem{fl1}R. Friedberg and T.D. Lee, Phys. Rev. D \textbf{15}, 1694, (1977).
\bibitem{fl2}R. Friedberg and T.D. Lee, Phys. Rev. D \textbf{16}, 1096, (1977).
\bibitem{fl3}R. Friedberg and T.D. Lee, Phys. Rev. D \textbf{18}, 1978, (1978).

\bibitem{QMDD1}G.N. Fowler, S. Raha, and R.M. Weiner, Z. Phys. C \textbf{9}, 271 (1981)
\bibitem{QMDD2} S. Chakrabarty, S. Raha, and B. Sinha, Phys. Lett. B\textbf{229}, 112 (1989).
\bibitem{QMDD3} O. G. Benrenuto and G. Lugones, Phys. Rev. D \textbf{51},  1989 (1995);
G. Lugones and O. G. Benrenuto, Phys. Rev. D \textbf{52},  1276 (1995).

\bibitem{QMC1} P. A. M. Guichon, Phys. Lett. B\textbf{200}, 235 (1988).
\bibitem{QMC2} K. Saito and A. W. Thomas, Phys. Lett. B\textbf{327}, 9 (1994).
\bibitem{QMC3} K. Saito, K. Tsushima and A. W. Thomas, Prog. Part. Nucl. Phys. \textbf{58}, 1 (2007), and references therein.

\bibitem{strange mesons} J. Schaffner, et al., Ann. Phys. (NY),  \textbf{235}, 35 (1994).

\bibitem{parameters} R.J. Furnstahl, B.D. Serot and H.B. Tang, Nucl. Phys. A \textbf{615}, 441 (1997).

\bibitem{OZI} C.B. Dover and A. Gal, Prog. Part. Nucl. Phys., Vol. 12 (Pergamon, Oxford, 1984);
B.K. Jinnings, Phys. Lett. B \textbf{246}, 325 (1990).

\bibitem{Bouyssy} A. Bouyssy, Nucl. Phys. A \textbf{290}, 429 (1977).
\bibitem{Hausmann} R. Hausmann and W. Weise, Nucl. Phys. A \textbf{491}, 601 (1989).

\bibitem{-18MeV} T. Fukuda, A. Higashi, Y. Matsuyama, C. Nagoshi, J. Nakao, M. Sekimoto, P. Tlusty, J.K. Ahn, et al.  Phys. Rev. C \textbf{58}, 1306 (1998).

\bibitem{Schaffner} J. Schaffner, et.al, Phys. Rev. Lett. \textbf{71}, 1328 (1993);
J. Schaffner, I. Mishustin, Phys. Rev. C \textbf{53}, 1416 (1996).

\bibitem{Y-Y old 1} M. Danysz et al., Nucl. Phys.  \textbf{49}, 121 (1963);
R.H. Dalitz, D.H. Davis, P.H. Fowler, A. Montwill, J.Pniewski, and J.A. Zakrzewki, Proc. R. Soc. London, Ser. A \textbf{426}, 1 (1989).

\bibitem{Y-Y old 2} D.J. Prowse,  Phys. Rev. Lett. \textbf{17}, 782 (1966).

\bibitem{Y-Y old 3} S. Aoki et al., Prog. Theor. Phys. \textbf{85}, 1287 (1991);
C.B. Dover et al., Phys. Rev. C   \textbf{44}, 1905 (1991).

\bibitem{Y-Y new} H. Takahashi et al., Phys. Rev. Lett.   \textbf{87}, 212502 (2001).

\bibitem{Y-Y 2010} K. Nakazawa, Nucl. Phys. A \textbf{835}, 207 (2010).

\bibitem{Multi-lambda1} M. Barranco, R. J. Lombard, S. Marcos and S. A. Moszkowski,  Phys. Rev. C \textbf{44}, 178 (1991).
\bibitem{Multi-lambda2} L. L. Zhang, H. Q. Song  and R. K. Su, J. Phys. G \textbf{23}, 557, (1997).
\bibitem{Multi-lambda3} H.-J. Schulze, M. Baldo, U. Lombardo, J. Cugnon and A. Lejeune, Phys. Rev. C \textbf{57}, 704 (1998).

\end{references}
\end{document}